\documentclass[12pt]{article}
\usepackage{amssymb, graphicx}
\def\be{\begin{equation}}
\def\ee{\end{equation}}
\def\bes{\begin{eqnarray}}
\def\ees{\end{eqnarray}}
\def\p{\partial}

\def\half{{\textstyle{1\over2}}}

\catcode`\@=11
\def\@citex[#1]#2{%
\if@filesw \immediate \write \@auxout {\string \citation {#2}}\fi
\@tempcntb\m@ne \let\@h@ld\relax \def\@citea{}%
\@cite{%
  \@for \@citeb:=#2\do {%
    \@ifundefined {b@\@citeb}%
      {\@h@ld\@citea\@tempcntb\m@ne{\bf ?}%
      \@warning {Citation `\@citeb ' on page \thepage \space undefined}}%
      {\@tempcnta\@tempcntb \advance\@tempcnta\@ne%
      \@tempcntb\number\csname b@\@citeb \endcsname \relax%
      \ifnum\@tempcnta=\@tempcntb 
        \ifx\@h@ld\relax%
          \edef \@h@ld{\@citea\csname b@\@citeb\endcsname}%
        \else%
          \edef\@h@ld{\ifmmode{-}\else--\fi\csname b@\@citeb\endcsname}%
        \fi%
      \else
        \@h@ld\@citea\csname b@\@citeb \endcsname%
        \let\@h@ld\relax%
      \fi}%
    \def\@citea{,\penalty\@highpenalty\,}%
  }\@h@ld
}{#1}}

\def\@citeb#1#2{{[#1]\if@tempswa , #2\fi}}
\def\@citeu#1#2{{$^{#1}$\if@tempswa , #2\fi }}
\def\@citep#1#2{{#1\if@tempswa , #2\fi}}

\def\bcites{         
        \catcode`\@=11
        \let\@cite=\@citeb
        \catcode`\@=12
}

\def\upcites{         
        \catcode`\@=11
        \let\@cite=\@citeu
        \catcode`\@=12
}

\def\plaincites{      
        \catcode`\@=11
        \let\@cite=\@citep
        \catcode`\@=12
}

\newcount\hour
\newcount\minute
\newtoks\amorpm
\hour=\time\divide\hour by 60
\minute=\time{\multiply\hour by 60 \global\advance\minute by-\hour}
\edef\standardtime{{\ifnum\hour<12 \global\amorpm={am}%
        \else\global\amorpm={pm}\advance\hour by-12 \fi
        \ifnum\hour=0 \hour=12 \fi
        \number\hour:\ifnum\minute<10 0\fi\number\minute\the\amorpm}}
\edef\militarytime{\number\hour:\ifnum\minute<10 0\fi\number\minute}

\def\draftlabel#1{{\@bsphack\if@filesw {\let\thepage\relax
   \xdef\@gtempa{\write\@auxout{\string
      \newlabel{#1}{{\@currentlabel}{\thepage}}}}}\@gtempa
   \if@nobreak \ifvmode\nobreak\fi\fi\fi\@esphack}
        \gdef\@eqnlabel{#1}}
\def\@eqnlabel{}
\def\@vacuum{}
\def\marginnote#1{}
\def\draftmarginnote#1{\marginpar{\raggedright\scriptsize\tt#1}}
\def\draft{
        \pagestyle{plain}
        \overfullrule=2pt
        \oddsidemargin -.5truein
        \def\@oddhead{\sl \phantom{\today\quad\militarytime} \hfil
        \smash{\Large\sl DRAFT} \hfil \today\quad\militarytime}
        \let\@evenhead\@oddhead
        \let\label=\draftlabel
        \let\marginnote=\draftmarginnote
        \def\ps@empty{\let\@mkboth\@gobbletwo
        \def\@oddfoot{\hfil \smash{\Large\sl DRAFT} \hfil}
        \let\@evenfoot\@oddhead}
        \def\@eqnnum{(\theequation)\rlap{\kern\marginparsep\tt\@eqnlabel}%
        \global\let\@eqnlabel\@vacuum}  }

\begin{document}

\hfill UTHET-04-1101

\begin{center}
\Large
{\bf Constrained Perturbative Expansion of the DGP Model\footnote{Research supported in part by the DoE under grant DE-FG05-91ER40627.}}
\normalsize

\vspace{1.8cm}

{\bf Chad Middleton\footnote{E-mail: {\sf cmiddle1@utk.edu}} and George 
Siopsis\footnote{E-mail: {\sf siopsis@tennessee.edu}}}\\ Department of Physics
and Astronomy, \\
The University of Tennessee, Knoxville, \\
TN 37996 - 1200, USA.
\end{center}

\vspace{1.8cm}

\centerline{\bf Abstract}

We address the vDVZ discontinuity of the 5D DGP model which consists of a
3-brane residing in a flat, infinite-volume bulk.
Following a suggestion by Gabadadze {\tt [hep-th/0403161]},
we implement a constrained perturbative expansion parametrized by brane gauge
parameters.
We explore the parameter space and show that the DGP solution exhibiting the
vDVZ discontinuity corresponds to a set of measure zero.

\newpage
The weakness of the gravitational force has been successfully explained by
postulating the existence of extra dimensions \cite{WEAK}.  The effect of the 
extra dimensions is a high-energy modification of Newton's Law of gravity due to
the tower of Kaluza-Klein modes.  When the extra dimensions are of infinite 
volume, light Kaluza-Klein modes may dominate even at low energies 
\cite{DGP,nDGP,Car}, therefore offering an attractive alternative to dark energy
for solving the cosmological constant problem.  Thus, unlike with finite-volume
extra space, $4D$ gravity is modified at astronomically large distances 
\cite{LDM}.\footnote {See \cite{Kyae} for a slightly different treatment which yields a $4D$ tensor structure at astronomical distances.}

The DGP model of a 3-brane residing in a $5D$ bulk of vanishing cosmological constant
\cite{DGP} is a ghost-free, general covariant theory where the $5D$ graviton mimics a $4D$ massive graviton on the brane.  The model appears to be plagued by a van Dam-Veltman-Zakharov (vDVZ) discontinuity \cite{vanDam,Zak}
 and has attracted much attention \cite{DISC,Cross,Gab}.    In 
refs.\cite{Cross}, solutions were found interpolating between regimes far from 
and near the Schwarzschild radius by keeping higher-order terms in the 
perturbative expansion.  It was thus shown that in the decoupling limit, one
recovers the standard four-dimensional, weak-field Schwarzschild metric.  

As has been recently argued in \cite{Gab,Por1} for the specific case of $D=5$,
the breakdown of the perturbative expansion at linear order is 
an artifact of the weak-field expansion itself and can be healed by adopting a {\it constrained} perturbative expansion.  Thus, instead of the incorporation of
higher-order terms into the linearized treatment, the
theory is regulated by a modification of the linearized theory itself.
After fixing the gauge in the bulk, a residual four-dimensional gauge invariance remains on the brane.
The graviton propagator is then rendered invertible by the addition of a term
in the action which would amount to a gauge-fixing term in four-dimensional
gravity.

Here, we present a generalized procedure of \cite{Gab}. We introduce a two-parameter family of gauge-fixing terms on the brane.
In the decoupling limit they amount to ordinary gauge-fixing terms and no
physical quantities depend on these gauge parameters.
We also fix the gauge in the bulk in terms of arbitrary gauge parameters.
In the absence of a brane, five-dimensional gauge-invariance guarantees that
no physical quantities will depend on the bulk gauge parameters.
We then explore the physical effects of these parameters away from the two
extremal limits (decoupling and absence of a brane).
We find that the graviton propagator in general has a well-defined decoupling
limit implying the absence of a vDVZ discontinuity.  The graviton propagator exhibits the expected crossover behavior and is found to be free of tachyonic asymptotic states.
The DGP solution~\cite{DGP} corresponds to a set of measure zero in our parameter space.

The DGP model describes a 3-brane on the boundary of a five-dimensional bulk-space $\Sigma$.  The action is
\be\label{DGP}
S_{DGP}=M^3\int_\Sigma d^4x dy\;\sqrt{-G}\;\mathcal R^{(5)}+\overline M^2\int_{\p \Sigma} d^4x \sqrt{-g}\;\mathcal R^{(4)}
\ee
where $\mathcal{R}^{(5)},(\mathcal{R}^{(4)})$ is the five- (four-) dimensional Ricci scalar.  We adopt the standard conventions $\eta_{AB}=\mbox{diag}[+----]\;;\;A,B=0,\dots,3,y\;;\;\mu\nu=0,\dots,3\;;\;i,j=1,2,3$.

Upon varying (\ref{DGP}), one arrives at the DGP field equations, which are
\be\label{DGPfeq}
M^3G_{AB}^{(5)}+\overline M^2G_{\mu\nu}^{(4)}\delta^\mu_A\delta^\nu_B\delta(y)=T_{\mu\nu}
\delta^\mu_A\delta^\nu_B\delta(y)
\ee
with the linearized solution given by
\bes\label{5DGPsol}
\tilde h_{\mu\nu}(p,y)&=&\frac{1}{\overline M^2(p^2+2m_bp)}
\left\{\tilde T_{\mu\nu}-\frac{1}{3}\left(\eta_{\mu\nu}+\frac{p_\mu p_\nu}{2m_bp}\right)\tilde T\right\}e^{-py}\nonumber\\
\tilde h_\alpha^\alpha&=&-\frac{\tilde T}{6\overline M^2m_bp}e^{-py}=\tilde h_y^y
\ees
which is written in terms of the $4D$ Euclidean momentum and graviton mass
\bes
p^2&=&-p^\mu p_\mu=-p_0^2+p_i^2=p_4^2+p_i^2\nonumber\\
m_b&=&M^3/\overline M^2
\ees
The solution bares a striking 
resemblance to that of massive gravity where the factor of $1/3$ instead of the Einstein 
factor of $1/2$ signals the existence of a vDVZ discontinuity. 
In the decoupling limit $(m_b\rightarrow 0)$, 4D Einstein gravity is not recovered and we do not obtain sensible dynamics for the longitudinal term with tensor structure of the 
form $p_\mu p_\nu$.  Although the $p_\mu p_\nu$ term does not contribute at linear level,
 it does enter nonlinear diagrams.

Generalizing \cite{Gab}, we define a {\it Constrained DGP Action} of the form
\be\label{ScDGP}
S_{cDGP}=S_{DGP}+S^{(5)}+ S^{(4)}
\ee
where $S_{DGP}$ is the DGP action given by (\ref{DGP}) and $S^{(4)}$ and $S^{(5)}$ are gauge-fixing terms in the decoupling limit ($m_b\to 0$) and absence of brane ($m_b\to\infty$), respectively.
Away from these two limits ($M,\overline M\neq 0$), these additional terms no longer simply fix the gauge; they alter the boundary conditions.

We start in the bulk by defining $S^{(5)}$ as follows
\be\label{Sgauge}
S^{(5)}=M^3\int_\Sigma d^4xdy\;\sqrt{-G}\left[\frac{B_5^2}{2\gamma}+\frac{B_\mu^2}{2\alpha}\right]
\ee
with
\bes\label{bugauge}
B_\mu&\equiv& \p_\mu h_y^y+a \p_\mu h_\alpha^\alpha-b\p^\alpha h_{\alpha\mu} \nonumber\\
B_y&\equiv&\p^\mu h_{\mu y}
\ees
where $\alpha,\gamma, a,b$ are arbitrary parameters on which no bulk physical
quantities should depend.
In the absence of the brane,  
eq.~(\ref{Sgauge}) amounts to standard gauge-fixing conditions. In general,
the $\alpha,\gamma\rightarrow 0$
limit should be taken at the end of the calculation to ensure that
\bes\label{bugauge0}
B_\mu&\rightarrow& 0\nonumber\\
B_y&\rightarrow& 0
\ees
Next, we define the gauge-fixing term $S^{(4)}$ on the brane.  For a brane of finite thickness, 
additional terms can arise on the brane world-volume and can survive in
the limit of the brane thickness tending to zero.  In addition, we note that the boundary equations receive no contribution from 
eq.~(\ref{Sgauge}) and are invariant under the $4D$ transformations \cite{Gab}
\be
h_{\mu\nu}|_{y=0}\rightarrow h_{\mu\nu}+\p_\mu\zeta_\nu+\p_\nu\zeta_\mu|_{y=0}
\ee
indicating a residual gauge freedom.  With the above in mind, we choose an additional brane action contribution
\be\label{Sgen}
 S^{(4)}=\lambda\; \overline M^2\int_{\p \Sigma} d^4x\sqrt{-g}\; \mathcal{B}_\nu \mathcal{B}^\nu
\ee
where
\be\label{4DGauge}
 \mathcal{B}_\nu \equiv \partial^\mu h_{\mu\nu} + \xi \partial_\nu h_\alpha^\alpha 
\ee
and we assume $\lambda >0$.
These additional action contributions modify the DGP model by explicitly breaking the 
$4D$ and $5D$ coordinate invariance.
Adopting this modified DGP model, we next obtain and solve the field equations.
Varying (\ref{ScDGP}), expanding around a flat background, and Fourier transforming, the first-order Einstein equations are as follows.
In the bulk, the transverse component is
\be\label{55}
\tilde h_\alpha^\alpha-\frac{p^\alpha p^\beta}{p^2}\tilde h_{\alpha\beta}
-\frac{1}{\alpha}
\left(\tilde h_y^y+a\tilde h_\alpha^\alpha-b \frac{p^\alpha p^\beta}{p^2}\tilde h_{\alpha\beta}\right)=0
\ee
The mixed components are
\be\label{alpha5}
i\partial_y (p^\alpha \tilde h_{\alpha\mu}-p_\mu \tilde h_\alpha^\alpha)+p^2\left(\tilde h_{\mu y}-\frac{1}{p^2}p_\mu p^\alpha \tilde h_{\alpha y}\right)+\frac{1}{\gamma}p_\mu p^\alpha \tilde h_{\alpha y}=0
\ee
and the components parallel to the brane are
\be\label{munu} \mathcal{G}_{\mu\nu}^{(5)} = 0 \ee
where
\bes\label{oper}
\mathcal{G}_{\mu\nu}^{(5)} &=&(p^2 - \p_y\p^y)(\tilde h_{\mu\nu}-\eta_{\mu\nu}\tilde h_\alpha^\alpha) -p_\mu p^\alpha \tilde h_{\alpha \nu}-p_\nu p^\alpha\tilde h_{\alpha \mu} +p_\mu p_\nu (\tilde h_\alpha^\alpha +\tilde h_y^y)-\eta_{\mu\nu}
(p^2\tilde h_y^y- p^\alpha p^\beta \tilde h_{\alpha\beta} )\nonumber\\
&+&\frac{1}{\alpha}\left[-b(p_\mu p_\nu \tilde h_y^y+a p_\mu p_\nu \tilde h_\alpha^\alpha-b p_\nu p^\alpha \tilde h_{\alpha\mu})+a\eta_{\mu\nu} (p^2\tilde h_y^y+a p^2\tilde h_\alpha^\alpha-b p^\alpha p^\beta\tilde h_{\alpha\beta} )\right]\nonumber\\
&+&i\p^y ( p_\nu\tilde h_{\mu y}+p_\mu\tilde h_{\nu y}-2\eta_{\mu\nu}p^\alpha \tilde h_{\alpha y} )
\ees
From eqs.~(\ref{55}) and (\ref{alpha5}), we obtain
\bes\label{sol5}
\tilde h_{\mu y}&=&\gamma \frac{p_\mu}{p^2}\; i\p_y \left(\tilde h_\alpha^\alpha-\frac{p^\alpha p^\beta}{p^2}\tilde h_{\alpha\beta}\right)\nonumber\\
\tilde h_y^y&=&b\frac{p^\alpha p^\beta}{p^2}\tilde h_{\alpha\beta}-a\tilde h_\alpha^\alpha+\alpha\left(\tilde h_\alpha^\alpha-\frac{p^\alpha p^\beta}{p^2}\tilde h_{\alpha\beta}\right)
\ees
Plugging these expressions into (\ref{oper}) and assuming the solution is of the form
\be \tilde h_{AB} (p,y) = \tilde h_{AB} (p) e^{-py} \ee
we may write $\mathcal{G}_{\mu\nu}^{(5)}$ entirely in terms of the $4D$ metric perturbations.
Dotting with the momentum, we obtain
\be p^\mu p^\nu \mathcal{G}_{\mu\nu}^{(5)} = (1+a-b) p^2 (p^2 \tilde h_\alpha^\alpha
- p^\alpha p^\beta \tilde h_{\alpha\beta} ) \ee
implying the constraint on the parameters
\be 1+a-b = 0 \ee
The vanishing of the divergence, $p^\nu \mathcal{G}_{\mu\nu}^{(5)} = 0$,
then implies
\be\label{eqnocon} p^\nu \tilde h_{\mu\nu} = p_\mu \frac{p^\alpha p^\beta}{p^2} \tilde h_{\alpha\beta} \ee
This is not an additional constraint on the metric. On general grounds, one may argue that
$p^\nu \tilde h_{\mu\nu} \propto p_\mu$, hence~(\ref{eqnocon}).
Using these results, we arrive at the expression
\be \mathcal{G}_{\mu\nu}^{(5)} = (\alpha -2\gamma -2a) (p_\mu p_\nu - \eta_{\mu\nu} p^2 ) \left(\tilde h_\alpha^\alpha-\frac{p^\alpha p^\beta}{p^2}\tilde h_{\alpha\beta}\right)
\ee
leading to a second constraint on the parameters,
\be \alpha -2\gamma -2a = 0 \ee
At the boundary, the Israel junction condition at $y=0$ yields
\be\label{munu1} \overline M^2 \mathcal{G}_{\mu\nu}^{(4)}
= \tilde T_{\mu\nu} \ee
where
\bes\label{oper1}
\mathcal{G}_{\mu\nu}^{(4)} &=& (p^2+2m_bp)(\tilde h_{\mu\nu}-\eta_{\mu\nu}\tilde h_\alpha^\alpha) -(1-\lambda)(p_\mu p^\alpha \tilde h_{\alpha \nu}+p_\nu p^\alpha\tilde h_{\alpha \mu})\nonumber\\
&+& (1+2\lambda\xi)(p_\mu p_\nu
\tilde h_\alpha^\alpha+\eta_{\mu\nu}p^\alpha p^\beta \tilde h_{\alpha\beta})+2\lambda\xi^2\eta_{\mu\nu}p^2\tilde h_\alpha^\alpha \nonumber\\
&+&2\gamma m_bp\left( \eta_{\mu\nu} - \frac{p_\mu p_\nu}{p^2} \right)\left(\tilde h_\alpha^\alpha-\frac{p^\alpha p^\beta}{p^2}\tilde h_{\alpha\beta}\right)
\ees
Eq.~(\ref{munu1}) can be solved for arbitrary parameters $\lambda,\xi$ and $\gamma$.  We obtain on the brane
\be\label{sol}
\tilde h_{\mu\nu}(p)=\frac{1}{\overline M^2(p^2+2m_bp)}\left\{\tilde T_{\mu\nu}-\left(\eta_{\mu\nu}\mathcal{C}_1+\frac{p_\mu p_\nu}{p^2}\mathcal{C}_2\right)\tilde T\right\}
\ee
where
\bes
\mathcal{C}_1 &=&\frac{2m_b^2+2\lambda(1+\xi)(1-\xi +2\gamma(1+\xi))m_bp+\lambda(1+\xi)^2p^2}{6m_b^2+4\lambda(1+\xi)(1-2\xi+3\gamma (1+\xi))m_bp+2\lambda(1+\xi)^2p^2}\nonumber\\
\mathcal{C}_2 &=&\frac{(1-2\lambda(1+\xi)(1+2\gamma(1+\xi))m_bp-\lambda(1+\xi)(1+2\xi)p^2}{6m_b^2+4\lambda(1+\xi)(1-2\xi+3\gamma (1+\xi))m_bp+2\lambda(1+\xi)^2p^2}\nonumber\\
\ees
Notice that the $4D$ metric
perturbations, when convoluted with a conserved tensor $\tilde T^{\prime\mu\nu}$,
\be\label{gravprop}
 \tilde h_{\mu\nu} \tilde T^{\prime\mu\nu} = \frac{1}{\overline M^2(p^2+2m_bp)}\left\{\tilde T_{\mu\nu}\tilde T^{\prime\mu\nu} -\mathcal{C}_1\tilde T\tilde T'\right\}
\ee
are   
still dependent on the parameters $\lambda, \xi$ and $\gamma$.
Examining the $4D$ momentum dependence of the metric perturbations, we find
in the large momentum regime ($p\gg m_b$),
\be\label{eq4D}
\tilde h_{\mu\nu} \tilde T^{\prime\mu\nu} \simeq \frac{1}{\overline M^2p^2}\left\{\tilde T_{\mu\nu} \tilde T^{\prime\mu\nu} - \half
\tilde T\tilde T' \right\} \ee
recovering 4D Einstein gravity, and in the small momentum limit ($p\ll m_b$),
\be\label{eq5D}
\tilde h_{\mu\nu} \tilde T^{\prime\mu\nu} \simeq \frac{1}{2 M^3 p}\left\{ \tilde T_{\mu\nu} \tilde T^{\prime\mu\nu} - 
{\textstyle{\frac{1}{3}}}\tilde T \tilde T'\right\}
\ee
exhibiting 5D behavior, as expected.
Notice that in both limits, the transverse components of the metric on the brane are independent of the parameters $\lambda, \xi$ and $\gamma$.
In the intermediate range, the propagator smoothly switches from the 4D expression~(\ref{eq4D}) to the 5D expression~(\ref{eq5D}) as the momentum decreases.
This crossover behavior depends on the parameters $\lambda, \xi$ and $\gamma$.

In the decoupling limit, $m_b\rightarrow 0$, the graviton propagator yields the standard $4D$ Einstein solution on the brane demonstrating the absence of a vDVZ discontinuity.
This is the case in the entire parameter space except for a set of measure zero defined by
\be \xi = -1\ee
For this special choice, the parameters become true gauge parameters throughout
the entire range of momenta. We obtain
\bes\label{eqDGP}
\tilde h_{\mu\nu}(p,y)&=&\frac{1}{\overline M^2(p^2+2m_bp)}
\left\{\tilde T_{\mu\nu}-\frac{1}{3}\left(\eta_{\mu\nu}+\frac{p_\mu p_\nu}{2m_bp}\right)\tilde T\right\}e^{-py}\nonumber\\
\tilde h_y^y (p,y) &=& -\frac{1}{6\overline M^2m_bp}\ \tilde T \ e^{-py}\nonumber\\
\tilde h_{\mu y} (p,y) &=&0
\ees
which is independent of $\alpha,\gamma$. Also, the constraints $B_\mu=B_5=0$ for general $\alpha,\gamma$ showing that they represent gauge-fixing conditions.
This is the standard DGP model~\cite{DGP}.

It should also be noted that for the choice of parameters $\lambda=1,\; \xi=-1/2$, we recover the
model proposed by Gabadadze~\cite{Gab},
\bes
\tilde h_{\mu\nu}(p,y)&=&\frac{1}{\overline M^2(p^2+2m_bp)}
\left\{\tilde T_{\mu\nu}-\frac{1}{2}\eta_{\mu\nu}\frac{(p^2+4m_bp)}{(p^2+6m_bp)}\tilde T\right\}e^{-py}\nonumber\\
\tilde h_y^y&=&\frac{p^\alpha p^\beta}{p^2}\tilde h_{\alpha\beta}+(\half\alpha+\gamma)\left(\tilde h_\alpha^\alpha-\frac{p^\alpha p^\beta}{p^2}\tilde h_{\alpha\beta}\right)\nonumber\\
\tilde h_{\mu y}&=&-i\gamma\frac{p_\mu}{p}\left(\tilde h_\alpha^\alpha-\frac{p^\alpha p^\beta}{p^2}\tilde h_{\alpha\beta}\right)
\ees
in the $\alpha,\gamma\rightarrow 0$ limit.

We next wish to examine the poles of the propagator.  Taking the $\gamma\rightarrow 0$ limit, the transverse part of the propagator~(\ref{gravprop}) can be written in a form
explicitly revealing its pole structure,
\be
\tilde h_{\mu\nu}\tilde T^{'\mu\nu}=\frac{1}{\overline M^2\, (p^2+2m_bp)}\tilde T_{\mu\nu}\tilde T^{'\mu\nu}-\frac{1}{3\overline M^2}\ \mathcal{C} (p) \tilde T\tilde T'
\ee
where
\be
\mathcal{C} (p) =\frac{1}{p^2+2m_bp}+\frac{1}{2(c_+-c_-)}\left[\frac{c_+}{(p^2+c_+m_bp)}-\frac{c_-}{(p^2+c_-m_bp)}\right]
\ee
The location of the poles is determined by the coefficients
\be
c_\pm=c_\pm(\xi,\lambda)=\frac{1}{1+\xi}\left[1-2\xi\pm\sqrt{(1-2\xi)^2-\frac{3}{\lambda}}\;\right]
\ee
For $p\gg m_b$, $\mathcal{C} (p) \approx \frac{3}{2p^2}$ and we recover the
4D expression~(\ref{eq4D}).
The poles are significant for momenta $p\lesssim m_b$.
As was shown in \cite{Gab}, the $p=-2m_b$ pole lies on the second Riemann sheet in the Minkowski four-momentum complex plane,
where $p^2=s\, \exp{(-i\pi)}$, $s=p_\mu p^\mu$.
This pole corresponds to a non-physical resonance and indicates an intermediate, metastable state.  
This can be seen from the $p=\pm\sqrt{-s}$ dependence of the propagator which indicates that the propagator is multi-valued and the complex $s$-plane has two sheets with a branch cut on the positive real axis.
For the choice of $p=\sqrt{-s}$, we obtain a non-physical resonance and a propagator which decays with the bulk coordinate.  



 
The other two poles are located at $p = -c_\pm m_b$ and depend on the parameters
$\xi$ and $\lambda$.
In the $(\xi,\lambda)$-plane,
above the curve
\be\label{eqcurve} \lambda = \frac{3}{(1-2\xi)^2} \ee
both poles lie on the negative real axis in the complex $s$-plane, since $c_\pm \in \mathbb{R}$.
Moreover, $c_\pm > 0$ for $-1<\xi < 1/2$. In this strip, the two poles
are in the second Riemann sheet (corresponding to the choice $p = \sqrt{-s}$)
and are thus unphysical.
In the special case $\xi=-1/2$, $\lambda = 1$, the pole at $p=-c_-m_b$
coincides with the pole at $p=-2m_b$; this is the Gabadadze model~\cite{Gab}.
As we approach the curve~(\ref{eqcurve}), the two poles come together.
Below the curve~(\ref{eqcurve}), $c_\pm$ become complex and $c_+ = c_-^*$.
In this case the poles are no longer on the real axis;
we obtain a resonance with a momentum independent decay width, in addition to the pole at $p=-2m_b$.

As $\xi\to -1$, the two poles $p=c_\pm m_b$ become infinite and $\mathcal{C}(p)
\to (p^2+2m_bp)^{-1}$.
This is a singular case; dependence on the $\lambda$ parameter disappears
and the propagator turns into the DGP expression~(\ref{eqDGP})~\cite{DGP} which is plagued by the vDVZ discontinuity.

To the left of $\xi = -1$ (as well as for $\xi >1/2$), both $c_\pm < 0$;
therefore, the poles $p= -c_\pm m_b$
are tachyons, signaling instability of the solution.
Were we to choose $p= -\sqrt{-s}$, instead, we would place these two poles on
the second Riemann sheet, but then the third pole at $p=-2m_b$ would turn into
a tachyon.

\begin{figure}
\centerline{\includegraphics[height=7in]{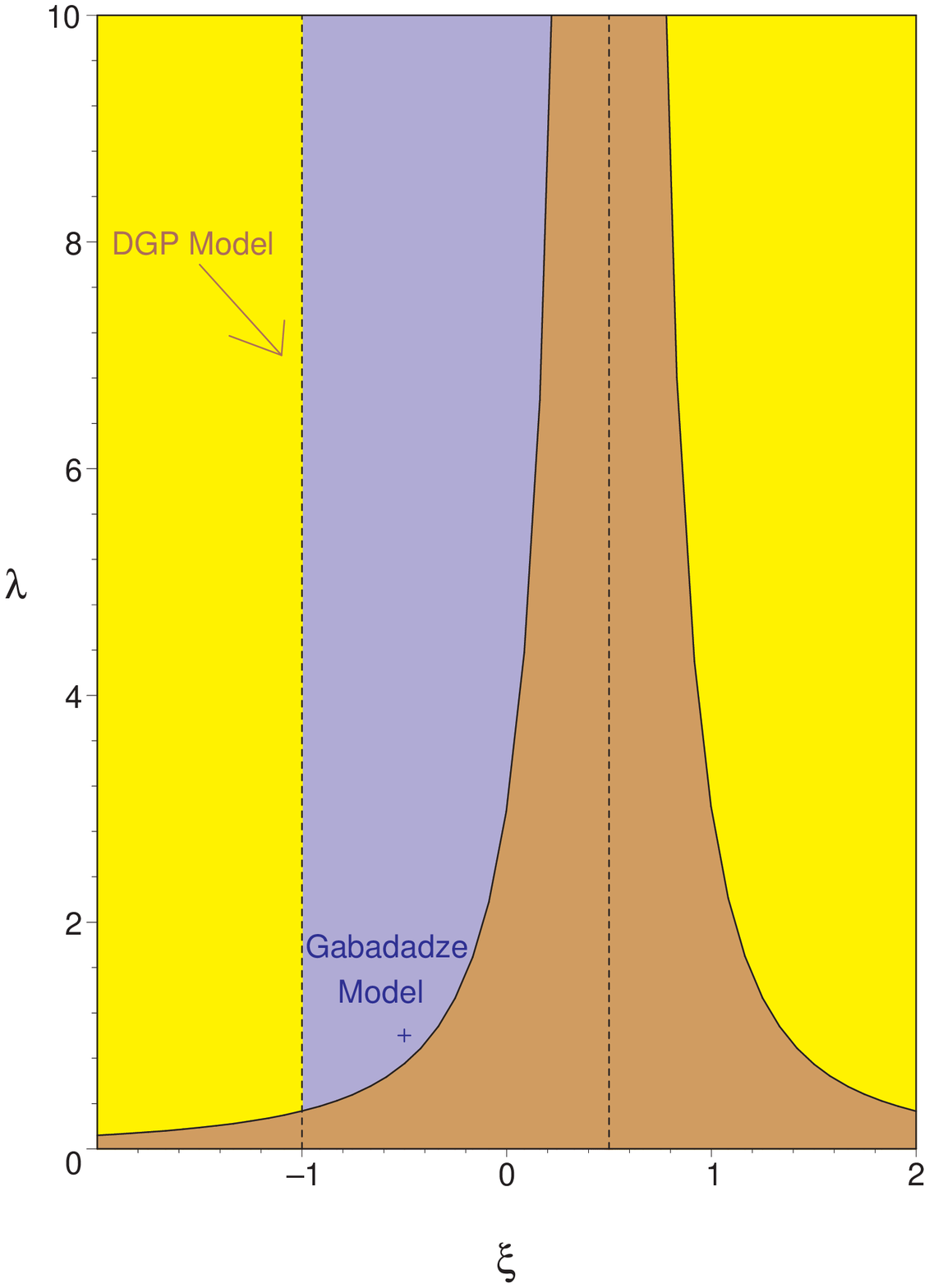}}
\caption{
\label{fig1}
The two-dimensional $(\xi,\lambda)$ parameter space.
Above the curve~(\ref{eqcurve}), all poles of the propagator are real.
Within the strip $-1<\xi < 1/2$, only unphysical resonances appear;
outside, we have tachyons (instability).
Below the curve~(\ref{eqcurve}), we have one real pole and a resonance with
momentum independent decay width.
The DGP model~\cite{DGP} is represented by the line $\xi = -1$;
the Gabadadze model~\cite{Gab} by the point $\xi = -1/2$, $\lambda = 1$.
}
\end{figure}

The above results are illustrated by the two-dimensional plot of
the $(\xi,\lambda)$ parameter space in Figure~\ref{fig1}.

In summary, we generalized the constrained perturbative model of \cite{Gab} and calculated the graviton propagator.
The first-order contribution to the perturbative expansion depended explicitly
on parameters which are gauge parameters in the bulk (in the absence of a brane)
and on the brane (in the decoupling limit), respectively.
These parameters determine the details of the distance-dependent, crossover behavior of the propagator and the position of the poles of the graviton propagator.
At low momenta, we obtained a 5D behavior whereas at high momenta we recovered
4D gravity demonstrating the absence of a vDVZ discontinuity.
In addition, we found a range of parameter values which yielded non-physical resonances corresponding to intermediate, metastable states.
For a special choice of parameters (representing a set of measure zero in the parameter space), we recovered the standard DGP model~\cite{DGP}.
This choice represented a set of measure zero in the parameter space which is plagued by the vDVZ discontinuity.

It would be desirable to understand the origin of these parameters better as physical quantities depend on them.
They may represent different physical setups (embeddings of the brane in the
bulk) or ``schemes'' (similar to QCD) which are artifacts of the perturbative expansion and would
be resolved once higher-order terms are included.
We hope to report on progress in this direction shortly.

\end{document}